\documentclass[openacc]{rsproca_new}

\usepackage[numbers]{natbib}
\usepackage{journalnames}

\begin{document}

\title{The upper atmospheres of Uranus and Neptune}

\author{
Henrik Melin$^{1}$} 

\address{$^{1}$School of Physics \& Astronomy, University of Leicester, UK}

\subject{Uranus, Neptune, Upper Atmosphere, H$_3^+$}

\keywords{Uranus, Neptune, Aeronomy}

\corres{Henrik Melin \\
\email{henrik.melin@leicester.ac.uk}}

\begin{abstract}
We review the current understanding of the upper atmospheres of Uranus and Neptune, and explore the upcoming opportunities available to study these exciting planets. The ice giants are the least understood planets in the solar system, having been only visited by a single spacecraft, in 1986 and 1989, respectively. The upper atmosphere plays a critical role in connecting the atmosphere to the forces and processes contained within the magnetic field. For example, auroral current systems can drive charged particles into the atmosphere, heating it by way of Joule heating. Ground-based observations of H$_3^+$ provides a powerful remote diagnostic of the physical properties and processes that occur within the upper atmosphere, and a rich data set exists for Uranus. These observations span almost three decades and have revealed that the upper atmosphere has continuously cooled between 1992 and 2018 at about 8 K/year, from $\sim$750 K to $\sim$500 K. The reason for this trend remain unclear, but could be related to seasonally driven changes in the Joule heating rates due to the tilted and offset magnetic field, or could be related to changing vertical distributions of hydrocarbons. H$_3^+$ has not yet been detected at Neptune, but this discovery provides low-hanging fruit for upcoming facilities such as the James Webb Space Telescope (JWST) and the next generation of 30 metre telescopes. Detecting H$_3^+$ at Neptune would enable the characterisation of its upper atmosphere for the first time since 1989. To fully understand the ice giants we need dedicated orbital missions, in the same way the Cassini spacecraft explored Saturn. Only by combining in-situ observations of the magnetic field with in-orbit remote sensing can we get the complete picture of how energy moves between the atmosphere and the magnetic field.

\end{abstract}



\begin{fmtext}
\end{fmtext}
\maketitle

\section{Introduction}

The upper atmosphere is the interface between the atmosphere below and the space environment beyond. Whilst there are substantial gaps in our understanding of this region at Uranus and Neptune, both planets offer completely unique environments compared to elsewhere in the solar system. Far away from the Sun, with complex magnetic fields \citep{1986Sci...233...85N,1989Sci...246.1473N}, having large axial tilts producing extreme seasons, and with observed upper atmospheric temperatures being inexplicably hot \citep[e.g.][]{2019RSPTA.37780408M}, the ice giants are exciting laboratories in which to examine how atmospheres exchange energy with the surrounding space environment. 


The Cassini mission to Saturn and the Galileo and Juno missions to Jupiter have shown that these literal journeys of discovery provide invaluable in-situ and remote sensing data that drive our deepening understanding our celestial neighbours. A dedicated in-orbit mission to either Uranus or Neptune (or both!) would truly transform the way we see these intriguing planets. As momentum builds for such a mission, we here review our limited understanding of the upper atmosphere of Uranus and Neptune, starting with an brief overview of the chemistry and processes important for this region, followed by a discussion of the giant planet {\it energy crisis} in the context of the ice giants. We then review the results obtained from the Voyager 2 flyby of Uranus and Neptune, in 1986 and 1989, respectively, and what we have learned since spacecraft encounter, primarily through ground-based observations H$_3^+$ emission. Finally, we discuss the future opportunities for studying the upper atmospheres of the ice giants, with new facilities and dedicated spacecraft missions.

\section{An Upper Atmosphere Primer}

The upper atmosphere is defined as the region in the atmosphere where molecular diffusion dominates. Here, species separate out with altitude according to their atomic or molecular weights, so that their vertical distribution is governed by their scale height. At the giant planets, heavier species are separated out very quickly at the bottom of this region, so that it is dominated at higher altitudes by atomic and molecular hydrogen. The upper atmosphere is bounded towards the lower atmosphere (mesosphere, stratosphere, and troposphere) by the {\it homopause}, the line at which molecular diffusion and eddy diffusion are equal. Eddy diffusion, or turbulent mixing, is the process where parcels of atmosphere move in bulk, producing a well-mixed atmosphere. The high altitude boundary of the upper atmosphere is the {\it exobase}, where the mean-free-path is equal to the scale height of atomic hydrogen. 

The upper atmosphere has two fundamental components, the neutral thermosphere and the charged particle ionosphere. The ionosphere is formed either by extreme ultraviolet (EUV) solar photoionisation or by impact ionisation by charged particles being driven into the atmosphere by the auroral process. Atomic hydrogen is ionised by both solar and particle impact to form H$^+$ and molecular hydrogen is converted to H$_2^+$ via these reactions:
\begin{equation}
H_2 + h\nu \longrightarrow H_2^+ + e^-
\label{solarion}
\end{equation}
\begin{equation}
H_2 + e^* \longrightarrow H_2^+ + e^- + e^-
\label{eion}
\end{equation}
\begin{equation}
H_2 (\nu \geq 4) + H^+ \longrightarrow H_2^+ + H
\label{hion}
\end{equation}
where $h\nu$ represents solar EUV photons, and $e^*$ are energetic precipitating electrons.  Equation \ref{solarion} produces H$_2^+$ on the dayside of the planet, whereas Equation \ref{eion} produces H$_2^+$ about the magnetic poles of the planet, peaking where the field-aligned currents are the strongest. Equation \ref{hion} is endothermic and therefore requires excited molecular hydrogen \citep{1973SSRv...14..460M}. Once formed, H$_2^+$ reacts very quickly with the ambient H$_2$ to produce H$_3^+$ via this exothermic reaction, releasing 1.7 eV of heat:
\begin{equation}
H_2^+ + H_2 \longrightarrow H_3^+ + H
\label{makeit}
\end{equation}
This makes H$^+$, H$_3^+$, and e$^-$ the dominant species the ionosphere of the giant planets, but out of these only H$_3^+$ emits an electromagnetic spectrum, observable in the near-infrared. The number of H$_3^+$ ions present in the upper atmosphere is a balance between the ionisation rate of molecular hydrogen, producing H$_3^+$ (Equation \ref{makeit}), and the loss mechanisms outlined above. Once formed, H$_3^+$ is thermalised to the surrounding neutral atmosphere and via remote sensing of the H$_3^+$ spectrum in the near-infrared we can determine the temperature of the upper atmosphere and the line-of-sight integrated H$_3^+$ density. This is achieved by fitting the observed spectrum to a theoretical spectrum, e.g. using the open source H$_3^+$ modelling and fitting software package\footnote{https://github.com/henrikmelin/h3ppy}. This makes observations of H$_3^+$ a really powerful remote diagnostic of the physical conditions and processes that occur within the ionosphere and the enveloping thermosphere. 

The life-time of H$_3^+$ is principally dominated by the electron density, since the molecular ion is destroyed via dissociative recombination with electrons: 
\begin{equation}
H_3^+ + e^- \longrightarrow H_2 + H
\end{equation}
\begin{equation}
H_3^+ + e^- \longrightarrow H + H + H
\end{equation}
Both H$^+$ and H$_3^+$ can also react with other heavier species, and the following reactions are extremely efficient for any species $X$ with proton affinity greater than molecular hydrogen: 
\begin{equation}
H_3^+ + X \longrightarrow XH^+ + H_2
\label{ionch4}
\end{equation}
\begin{equation}
H^+ + X \longrightarrow XH^+ 
\label{ionch4}
\end{equation}
This very effectively destroys H$^+$ and H$_3^+$ close to the homopause where hydrocarbons (e.g. methane, acetylene, and ethane) are present. Methane is present in the stratosphere because it is transported vertically from depth, and methane photochemistry drives the production of species such as acetylene and ethane \citep[e.g.][]{1990Icar...88..448B,1992JGR....9711681B}. The fact that H$^+$ and H$_3^+$ and hydrocarbons cannot coexist has two implications. Firstly, a planet with strong vertical eddy mixing in the stratosphere move hydrocarbons up to higher altitudes, reducing the size of the H$_3^+$ ionosphere. A change in the homopause height could therefore dramatically change the vertical profile of the ionosphere.  Secondly, if a planet has very energetic particle precipitation the ionisation peak may occur in the hydrocarbon layer below the homopause region. In this case, H$_3^+$ would be produced by the softer tail of the precipitating electron energy distribution, and observations of H$_3^+$ emission may provide an incomplete view of how the injection of auroral energy alters the state of the  atmosphere. 

Because the ionosphere is charged, its constituents are subject to the forces exerted by the magnetic field. This region is therefore an important interface layer between the planet and the magnetosphere. For example, the field lines that are accelerated away from rigid rotation in the magnetosphere will drag (or pull) ionospheric plasma, imparting momentum to the atmosphere via ion-neutral collisions. Conversely, the movement of ionospheric plasma can impose additional currents into the magnetospheric system, moving energy from the atmosphere into the wider magnetosphere. 

Auroral emissions are produced when field-aligned currents \citep[e.g.][]{2013JGRA..118.2897C}, forming part of the vast auroral current circuit that close in the ionosphere, interact with the atmosphere either via excitation or ionisation of its constituents. The former primarily produces emissions in the ultraviolet by H and H$_2$, whilst the latter produces H$_3^+$ (via Equations \ref{eion} and \ref{makeit}) which emits thermally in the near-infrared. When large scale magnetospheric currents close in the ionosphere they generate currents within the atmosphere, and heat it by way of Joule heating, which can produce up to TW of energy in the auroral regions of Jupiter and Saturn \citep{2004jpsm.book..185Y}. For more details on the aurora of ice giants, see {\it Lamy et al.} \cite{lamy2020} in this issue. 


Energy is lost from the upper atmosphere via conduction of heat down to lower altitudes and by radiative cooling by H$_3^+$. This molecule is a very efficient emitter and can remove significant amounts of energy from the atmosphere. The amount energy radiated by H$_3^+$ is driven exponentially by the ambient temperature and linearly by the H$_3^+$ density, so the cooling becomes more efficient in hotter and denser environments. For example, radiative cooling is very important Jupiter's auroral region \citep{1997Icar..127..379L}, whilst H$_3^+$ cooling is a relatively minor process at Saturn \cite{2007Icar..186..234M, 2013JPCA..117.9770M}. For Uranus and Neptune, the efficacy of this cooling mechanism is likely somewhere between Jupiter and Saturn.


\begin{figure}
\centering\includegraphics[width=5.1in]{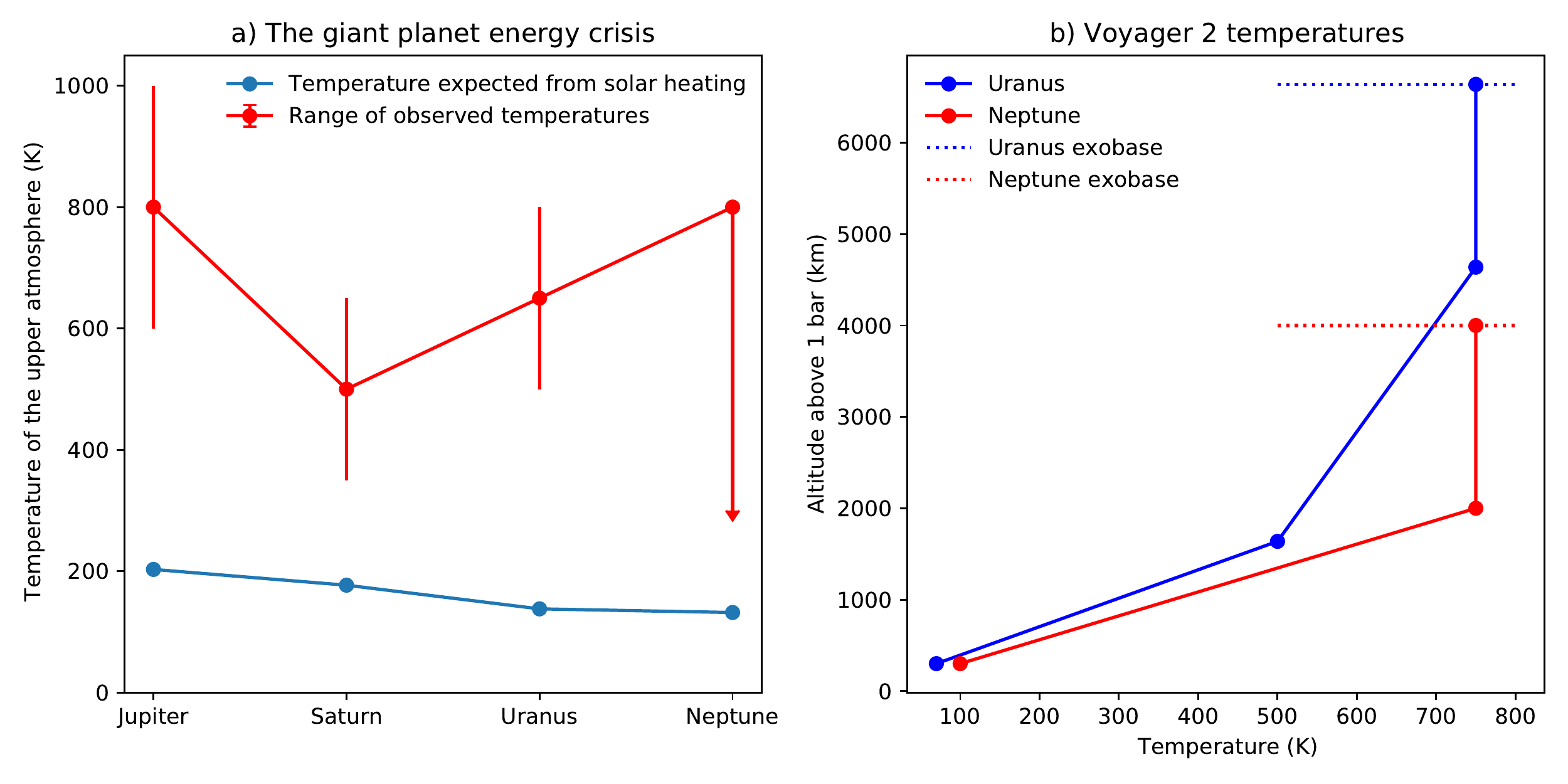}
\caption{a) The predicted and observed temperature of the upper atmospheres of the giant planets. There is only one temperature measurement for Neptune, but modelling suggests that the planet may have cooled since 1989, as indicated by the arrow \citep{moore2020} b) The vertical temperatures in the upper atmosphere of Uranus and Neptune derived from Voyager 2 UVS occultation observations \citep[adapted from][]{1989Sci...246.1459B,1986Sci...233...74B}. }
\label{figec}
\end{figure}

\section{The Giant Planet Energy Crisis}

The first measurements of the temperature of the upper atmospheres of the giant planets were derived from observations by the Voyager 1 \& 2 spacecrafts \cite{1979Sci...206..962S, 1981Sci...212..206B, 1986Sci...233...74B, 1989Sci...246.1459B}. They revealed a remarkable quandary: these temperatures are hotter by several hundreds of Kelvin than simple solar input models predict \citep{2004jpsm.book..185Y}. This is illustrated in Figure \ref{figec}a, showing the predicted versus the observed range of temperatures at non auroral latitudes for the four planets. These models do predict temperatures at the terrestrial planets that are in agreement with the observations -- what is the additional energy source at the giant planets that produce these high temperatures? This is known as the giant planet {\it energy crisis}, and presents a serious problem, we are fundamentally not understanding the atmospheres of half of the planets in our solar system -- and by extension, all of the giant planets in the universe.

Auroral heating can inject a huge amount of energy about the magnetic poles, but the observations  reveal the planets to be much hotter than solar models at {\it all} latitudes. For Jupiter and Saturn, where the magnetic poles are about co-located with the rotational poles, the auroral heating would need to be re-distributed from the pole down to the equator. These planets are very large and they are very fast rotators, producing very strong Coriolis forces that effectively prohibit the bulk movement of atmospheric packets down to lower latitudes. At Saturn, {\it Müller-Wodarg et al.} \citep{2019GeoRL..46.2372M} suggested that momentum loading of the upper atmosphere by gravity waves, effectively reducing the rotation rate that the atmosphere experiences, thus reducing the Coriolis forces, could act as a mechanism that permits energy to re-distribute effectively on the planet. This has recently been supported by analysis of ultraviolet Cassini occultation observations \cite{2020NatAs.tmp...75B}. 

Gravity waves have also been suggested as a potential energy source that can heat the upper atmosphere of the giant planets. These waves were directly observed in the temperature profile derived from the Galileo entry probe at Jupiter \citep{1997Sci...276..108Y} -- the only probe to sample a giant planet atmosphere in-situ. The effectiveness of gravity waves as a energy source for the upper atmosphere remains somewhat unclear, however. At Jupiter, {\it O'Donoghue et al. }  \citep{2016Natur.536..190O} observed the temperature of the H$_3^+$ ionosphere right above Jupiter's Great Red Spot (GRS), the largest storm in the solar system and a likely source of strong gravity waves. They derived temperatures of the upper atmosphere of over 1,600 K, hotter than anywhere else on the planet. In this isolated case it seems like that gravity waves are sufficient to heat the atmosphere at low latitudes, but many question marks remain -- why is the heating so confided to a narrow region close to the GRS? How is this energy re-distributed? What is the global distribution of these heating sources? 

Uranus and Neptune are very important pieces that need to be understood in the context of the energy crisis. Both planets are far away from the Sun and solar energy input is minimal, yet their upper atmospheres are both observed to be hotter than Saturn. Uranus has essentially no internal heat, whereas Neptune's internal heat source is substantial \citep{1990Icar...84...12P}. Similarly, the stratosphere of Uranus has limited vertical mixing, whilst Neptune is subject to strong mixing \cite[see ][and {\it Moses et al., this issue}]{2018Icar..307..124M}. This results in the homopause being at pressures a factor of three times greater at Uranus compared to Neptune. This renders the upper atmosphere of Uranus very deep, and the one of Neptune more shallow, and the lack of vertical mixing in the lower atmosphere of Uranus may limit the generation of gravity waves.

\begin{figure}[t]
\centering\includegraphics[width=5.1in]{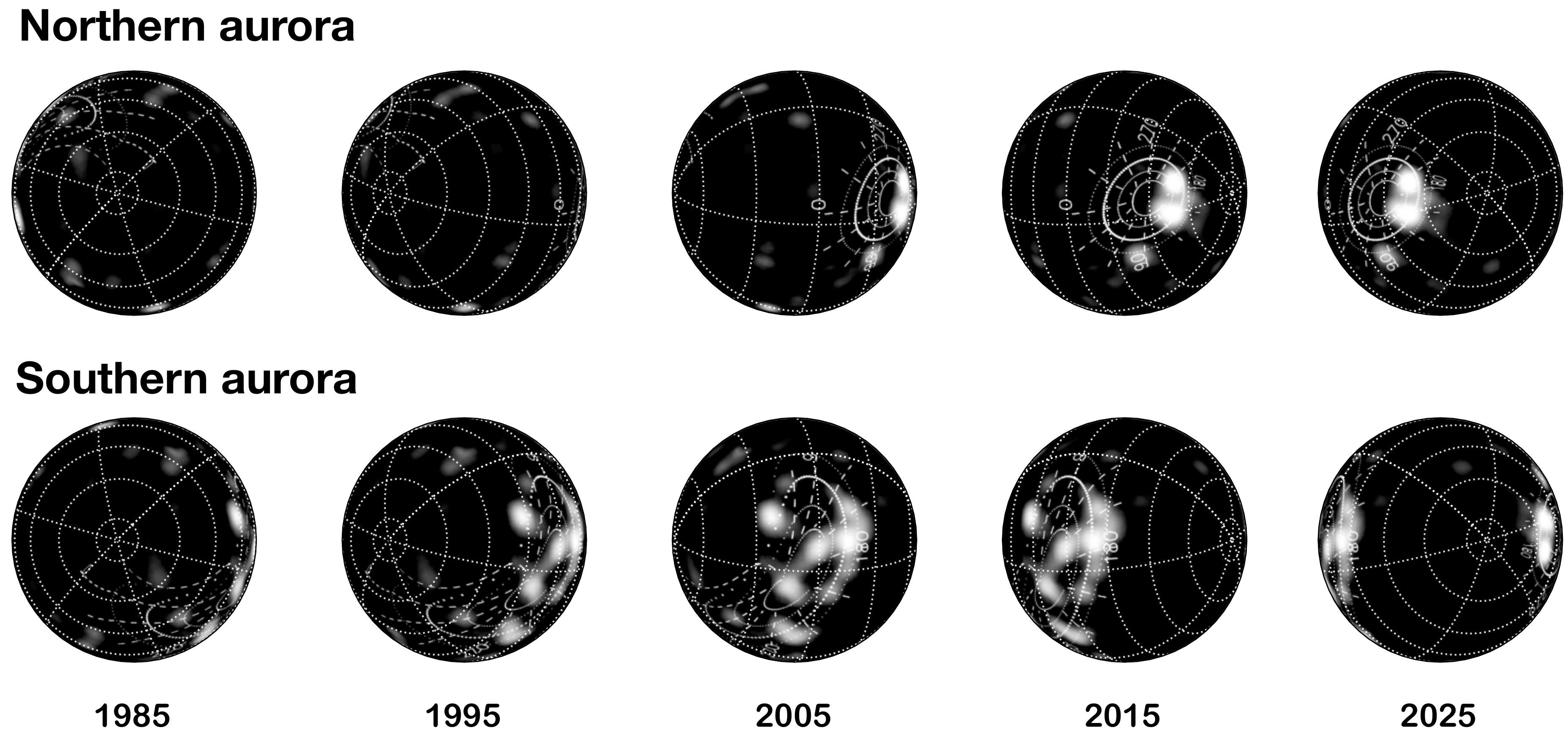}
\caption{The projected auroral observations of Uranus by {\it Herbert et al.} \cite{2009JGRA..11411206H} projected for different epochs, covering 40 years, as seen from the Earth. Light regions indicate bright auroral emission. The top row shows a rotational phase optimal for the northern auroral oval, and the bottom row is rotated 180 degrees longitude of that. The IAU north pole is oriented to the right. The peak brightness is 450 Rayleigh integrated between 95 to 110 nm. }
\label{fig_sim}
\end{figure}

Another important consideration for Uranus and Neptune is geometry. The magnetic field of Uranus is tilted from the rotational axis by 60 degrees, and is offset from the centre by 0.3 R$_U$. This places the magnetic poles close to the equator, as illustrated in Figure \ref{fig_sim}, which shows the auroral observations of {\it Herbert} \citep{2009JGRA..11411206H} projected onto different seasonal geometries of the planet, covering approximately half a Uranus year, between 1985 and 2025. The auroral emission and associated field-aligned currents are far away from the rotational poles, which limits the constraining effect of the Coriolis force on re-distributing auroral Joule heating. Similarly, the magnetic field of Neptune is tilted 47 degrees from the rotational axis, and is offset 0.55 R$_N$ from the centre of the planet. These geometries appear to render auroral Joule heating a more efficient means by which to globally heat the planet.




\section{The Voyager 2 flybys}

The Voyager 2 flybys of the ice giants laid the foundation of our understanding of these two systems -- please refer to the seminal {\it Uranus} book \citep{uranusbook} for more details. The spacecraft carries two instruments that can sample the upper atmosphere. Firstly, the Radio Science System \citep[RSS,][]{1977SSRv...21..207E} can broadcast a radio signal that is received at the Earth, and as the spacecraft moves behind the planet, the way the signal is attenuated is driven by the electron density in the ionosphere. The resulting vertical electron density profiles obtained by the spacecraft at both Uranus and Neptune showed large difference between dusk and dawn, high variability with altitude, and sharp layers below altitudes of 2,000 km \cite{1986Sci...233...79T, 1989Sci...246.1466T}. 

Secondly, the Ultraviolet Spectrograph \cite[UVS,][]{1977SSRv...21..183B} can both directly observe excited H and H$_2$ emissions from the upper atmosphere in the form of day-glow and auroral emissions, and perform occultation measurements, which yields vertical profiles of temperature and H and H$_2$ densities. The UVS instrument directly observed auroral emissions at Uranus \citep{2009JGRA..11411206H}. Figure \ref{fig_sim} shows this data projected for different geometries. These observations appear consistent with more recent ultraviolet observations obtained with the Hubble Space Telescope \citep[][and {\it Lamy et al.} in this issue]{2012GeoRL..39.7105L,2017JGRA..122.3997L,2018sf2a.conf...29L} and show emission in the form of discrete faint spots. At Neptune, no unambiguous detection of auroral emission made \citep{1989Sci...246.1459B} -- the only magnetised planet in our solar system without a single observation of its auroral emissions.

The only vertical temperature profiles of the upper atmospheres of Uranus and Neptune in existence were derived from UVS observations \cite{1989Sci...246.1459B,1986Sci...233...74B} -- shown in Figure \ref{figec}. Both planets were observed to have a temperature of 750 K at their respective exobase, but Uranus displayed a much more extended upper atmosphere compared to Neptune. This is because the weak vertical mixing lowers the altitude of the homopause \citep{2018Icar..307..124M}, producing a more extended upper atmosphere. It is noteworthy that the temperature profiles in Figure \ref{figec}b are constrained by only a handful of measurements, highlighting the fact that Uranus and Neptune remains the most poorly characterised planets in the solar system.

\begin{figure}[t]
\centering\includegraphics[width=5.1in]{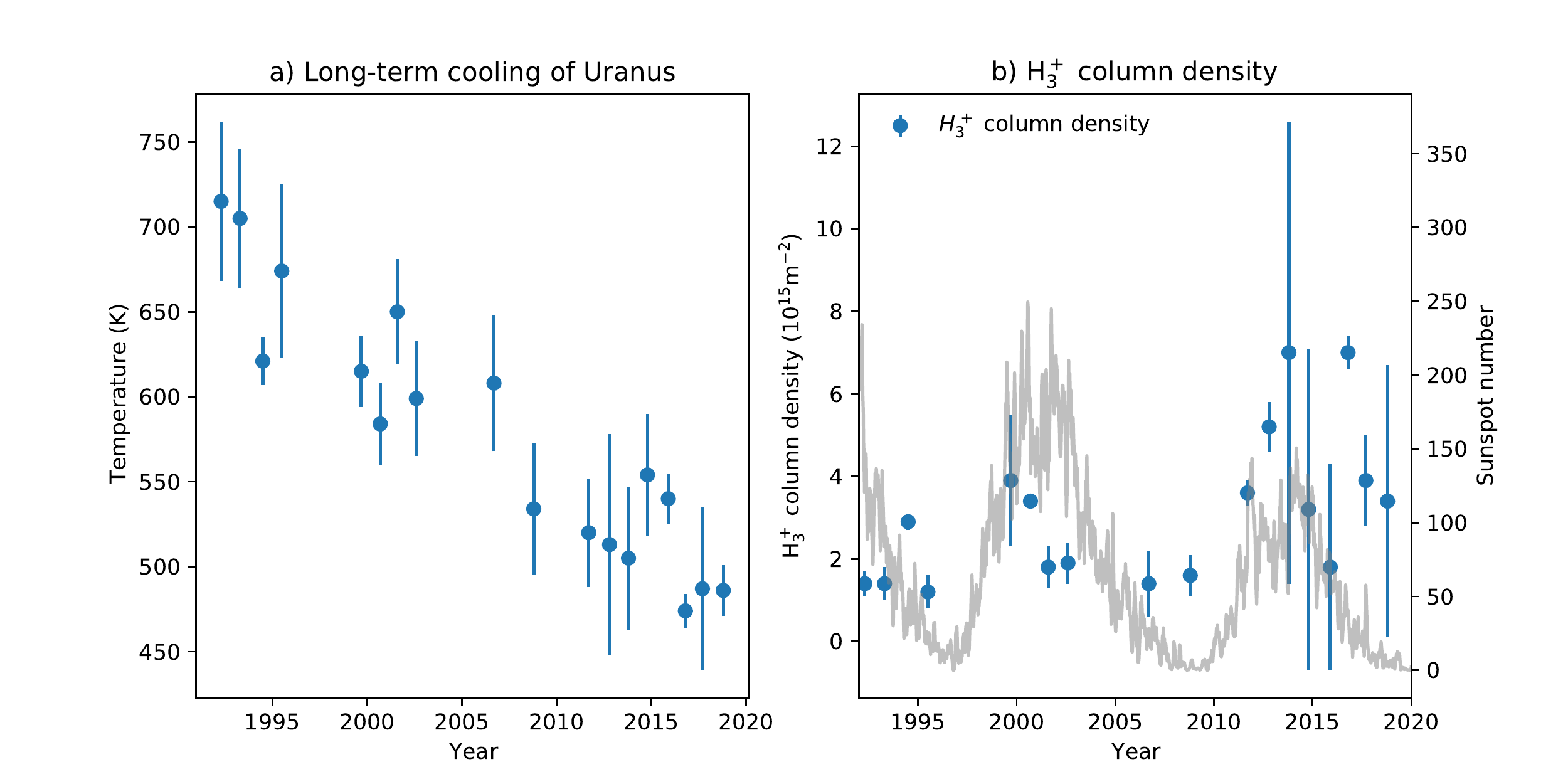}
\caption{a) The long-term cooling of the upper atmosphere of Uranus detailed in {\it Melin et al.} \cite{2013Icar..223..741M,2011ApJ...729..134M,2018MNRAS.474.3714M,2019RSPTA.37780408M}. The ionosphere has cooled by $\sim$8 K/year, consistently over two decades. b) The globally averaged H$_3^+$ column density for the same data-set, together with the monthly averaged sunspot number (grey, right axis), indicating the changing solar cycle. The data plotted here is listed in Table 3 in {\it Melin et al.} \citep{2019RSPTA.37780408M}
 } 
\label{longterm}
\end{figure}

\section{The Era of H$_3^+$}

Emissions from H$_3^+$ were discovered for the first time outside the laboratory at Jupiter in 1988 \citep{1989Natur.340..539D} as Voyager 2 was approaching Neptune. By analysing the observed H$_3^+$ near-infrared spectra we can derive the temperature of the upper atmosphere, the  column integrated density of the ionosphere, and radiative cooling rates. These products provide a powerful view through which we can begin to understand the processes that govern the upper atmosphere.

\subsection{Uranus}

H$_3^+$ was discovered by {\it Trafton et al.} \cite{1993ApJ...405..761T} in April 1992 using the United Kingdom Infrared Telescope (UKIRT), three months prior to the detection of H$_3^+$ of Saturn \citep{1993ApJ...408L.109G}. They observed a globally averaged temperature of 740$\pm$25 K, similar to the 750 K exospheric temperature derived by Voyager 2 (see Figure \ref{figec}). In many ways, this first detection paper became prophetic: they stated that the aim of future observations should be {\it "to determine the relative importance of the auroral and non-auroral processes"}. Almost three decades later, we still do not know how important the auroral process is in driving H$_3^+$ emission at Uranus. Observations obtained in 1993 showed that the variability on the observed intensity over four days was only 20\% \citep{1997ApJ...474L..73L}, suggesting that the solar ionisation mechanism for creating (Equation \ref{solarion}) H$_3^+$ dominates over particle impact ionisation (Equation \ref{eion}). 

{\it Encrenaz et al.} \citep{2003P&SS...51.1013E} analysed near-infrared observations from 2000 and 2001, placing them into context with previous observations. They suggested that the intensity of H$_3^+$ was a function of solar cycle, where the increased solar EUV flux at solar maximum would create a larger number of ions. This notion will be re-examined below, using all the available observations.

The first spatially resolved view of H$_3^+$ emission from Uranus was provided by \citep{1997ApJ...474L..73L}, with a very low S/N. An improved picture was provided by {\it Trafton et al.} \citep{1999ApJ...524.1059T}, who observed both quadrupole H$_2$ at 2 $\mu$m and H$_3^+$ with a north-south slit across the disk of the planet. The distribution of the two species were observed to be different, with H$_3^+$ peaking at the sub-solar point, whilst H$_2$ exhibited distinct brightening towards the limb of the disk of Uranus, consistent with an extended thermosphere, creating a large line of sight brightening at the limb. A sub-solar enhancement of H$_3^+$ is consistent with the idea that the solar process dominates the production of H$_3^+$. 

{\it Melin et al.} \citep{2011ApJ...729..134M} returned to all the available historical observations of H$_3^+$ from Uranus and discovered a long-term cooling trend in the globally averaged temperatures derived from fitting spectra. This cooling was interpreted as seasonal in nature, with solstice being warm and equinox being cool. Of course, characterising the temperature as Uranus moved away from equinox in 2007 and towards the 2028 solstice becomes critical for the interpretation of this long-term cooling, and as a result a concerted observing programme was developed in 2010 with the aim of observing Uranus multiple times per year for the next decade, primarily using the NASA Infrared Telescope Facility (IRTF), with the SpeX \citep{2003PASP..115..362R} and iSHELL \citep{2012SPIE.8446E..2CR} instruments. {\it Melin et al.} \cite{2013Icar..223..741M, 2018MNRAS.474.3714M,2019RSPTA.37780408M} performed and analysed these observations and the globally averaged temperature is shown in Figure \ref{longterm}a, and the column integrated H$_3^+$ density is shown in Figure \ref{longterm}b. In these figures, the error bars indicate the standard deviation of the measurements if multiple observations where performed in any one year, and the standard error if only one set of measurements were obtained. The temperature trend in Figure \ref{longterm}a is remarkable: Uranus has consistently cooled between 1992 and 2018, and no reversal in temperature was  observed at the equinox in 2007, which would be expected if the temperature variability was purely seasonal in nature. It remains unclear what could be causing this cooling, but Melin et al. \citep{2019RSPTA.37780408M} suggested that this could be driven by changes in the auroral Joule heating as the planet revolves about the Sun. Since the magnetic field is highly tilted away from the rotational axis and offset from the centre of the planet (see Figure \ref{fig_sim}), the upper atmosphere could potentially be subject to different Joule heating rates at every point in its orbit, perhaps rendering Uranus hot at one solstice and cold at the other.


Here, we put forward an alternative theory which might explain the long-term cooling in the upper atmosphere of Uranus, without the need for changing the vertical temperature profile. Equation \ref{ionch4} describes how H$_3^+$ ions are destroyed when they come in contact with hydrocarbons at homopause -- the base of the upper atmosphere. Therefore, an increase in the altitude of the homopause would reduce the density of H$_3^+$ and increase the observed temperature, since the remaining H$_3^+$ population would sample higher altitudes, where it is hotter (see Figure \ref{figec}). Conversely, a lower homopause would produce observations of higher densities and lower temperatures, as is the case in the present epoch (see Figure \ref{longterm}). In this manner, the temperature trend in Figure \ref{longterm}a can be achieved with a homopause altitude that changes with time, or a homopause altitude that is asymmetric across the planet and as the planet rotates, the observational geometry exposes different homopause altitudes towards the Earth. In this latter scenario there is no need to change any parameter, apart from geometry, to archived the observed long-term cooling trend. Whilst there is significant scatter in the H$_3^+$ density observations, they are broadly consistent with this notion, with higher densities being associated with lower temperatures. Changes in the homopause altitude could potentially be driven by changes to the global circulation pattern or by changes to the strength of the vertical mixing in the stratosphere. The feasibility of this mechanism will be explored with future modelling efforts.

Over the almost three decades since H$_3^+$ was discovered at Uranus, the advances in both telescope and instrument facilities have increased dramatically. Whilst our ability achieve greater and greater signal-to-noise for any given object in the night sky has increased many-fold, the long-term cooling of the upper atmosphere of Uranus means that the H$_3^+$ signal received at Earth is some 30 times fainter in this current epoch than it was in 1992, effectively offsetting the technological improvements. This makes Uranus a challenging object to observe, and if the current cooling trend continues, longer ground-based programmes are needed to determine the temperature of the planet.

Figure \ref{longterm}b also shows the sunspot number which is a direct proxy of the solar cycle. The density of H$_3^+$ is driven by the ionisation of molecular hydrogen, and so an increased solar EUV flux could potentially generate increased densities. Whilst high densities in Figure \ref{longterm}b do occur during solar maximum, they also occur during very low solar activity, and no direct trend can be inferred from the long-term data-set. The fact that the densities are larger towards 2010 onwards than the earlier measurements is of note, but could be related to an increased calibration uncertainty that can come with using instruments with narrower slits.

The HST observations of auroral emissions from Uranus \citep{2012GeoRL..39.7105L,2017JGRA..122.3997L,2018sf2a.conf...29L} show relatively weak auroral spots on the body of the planet, similar to Voyager 2 observed in 1986 (see Figure \ref{fig_sim}). {\it Lamy et al.} \cite{2012GeoRL..39.7105L} was the first to coordinate the observations with the arrival of a coronal mass ejection (CME) at the planet, which likely produced enhanced auroral emission. In 2017 {\it Lamy et al.} \citep{ 2018sf2a.conf...29L} performed a similar campagin, but this time adding ground-based imaging at H$_3^+$ wavelengths. Whilst H$_3^+$ auroral emissions were not detected, the northern (rotation) pole was observed to be brighter than the rest of the planet. Since bright emission can be driven by both temperature and density, determined via spectral analysis, it is currently unclear what drives this enhancement, but it could be indicative of a homopause at lower altitudes about the pole.


\subsection{The missing H$_3^+$ at Neptune}

H$_3^+$ emission remains undetected from Neptune. The first attempt to detect it was by \cite{1993ApJ...405..761T}, observing the planet for 58 minutes. In contrast, the latest attempt to detect the emission \citep{2018MNRAS.474.3714M} used 15.4 hours of exposure with the NASA IRTF iSHELL, producing an upper limit of the H$_3^+$ Q(1, 0$^-$) intensity of only 4\% compared to the first detection attempt. A comparison between derived upper limits of the intensity and density of H$_3^+$ is listed in Table \ref{neptunen}. {\it Moore et al.} \citep[][in this issue]{moore2020} explores in detail the potential reasons of why the ionosphere of Neptune is observed to be severely depleted.

\begin{table}[t]
\begin{tabular}{l|r|r}
& Upper limit for the & Upper limit for the \\
 & H$_3^+$ Q(1, 0$^-$) intensity & H$_3^+$ Q(1, 0$^-$) density  \\
Reference & (nWm$^{-2}$sr$^{-1}$) & at 550 K ($10^{13}$ m$^{-2}$)\\
\hline
Trafton et al. (1993) \cite{1993ApJ...405..761T} & 22 &  27  \\
Feuctgruber et al. (2003) \cite{2003AA...403L...7F} & 24  & 29   \\
Melin et al. (2011) \cite{2011MNRAS.410..641M} & 1.3 & 1.5 \\
Melin et al. (2018) \citep{2018MNRAS.474.3714M} &  0.8 & 1.0   \\
\end{tabular}
\caption{The observed upper limits for the H$_3^+$ Q(1, 0$^-)$ intensity and column density of Neptune. The column density calculation assumes a temperature of 550 K. For reference, the model of Lyons et al. \cite{1995Sci...267..648L} predicts a H$_3^+$ column density of $\sim 5 \times 10^{13}$ m$^{-3}$.\label{neptunen}}
\end{table}

\begin{figure}[t]
\centering\includegraphics[width=5.1in]{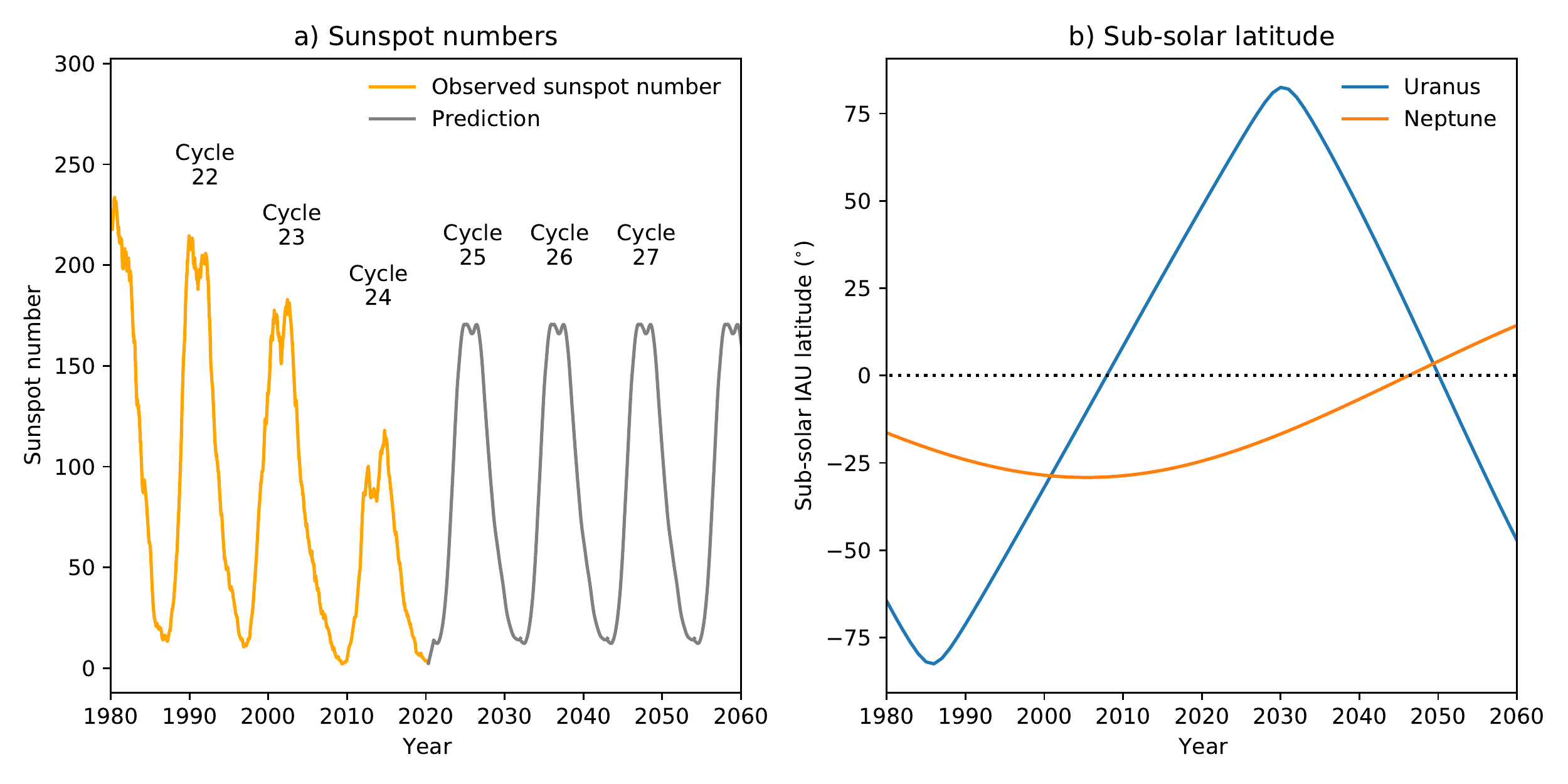}
\caption{a) The yearly averaged sunspot number, and a rudimentary prediction based on the average length of the solar cycle and the shape of the previous five cycles. b) The sub-solar latitude of Uranus and Neptune. In the late 2040s both of the planets will be in equinox conditions.  }
\label{future}
\end{figure}

\section{Future Opportunities}

There are a number of fundamental outstanding questions about the upper atmosphere of the ice giants, and future facilities such as the James Webb Space Telescope (JWST) and the next generation of 30 metre telescopes can provide unprecedented resolution and sensitivity to address these. There are already H$_3^+$ observations of Uranus planned as part of JWST's Guaranteed Time Observing programme, due to be obtained during  Cycle 1. 

Of course, the most detailed view of a distant planet can only be provided by an orbiting spacecraft. Because of the large distances involved, the transfer times are very long, and even if an ice giant mission concept was selected by a space agency tomorrow, it would likely not arrive until the 2040s [e.g. see {\it Fletcher et al., this issue}]

Figure \ref{future}a shows the observed sunspot number (orange line), a direct indicator of the solar activity along the solar cycle. The grey line shows a simplistic prediction of future solar cycles based on shape of the last five cycles, and the average solar cycle length of 11.04 years. It shows that a spacecraft arriving at the ice giants in the late 2040s may experience a solar maximum. Since CMEs are more common during solar maximum \cite{10.1007/978-94-009-4612-5_13}, this could be conducive to driving bright auroral emissions, making the ionosphere-magnetosphere system more dynamic in this epoch. Figure \ref{future}b shows the sub-solar latitude for both Uranus and Neptune betwen 1980 and 2060. A spacecraft arriving in the late 2040s would sample equinox conditions at both Uranus and Neptune. In the 1980s, Voyager 2 sampled Uranus at solstice and Neptune at an epoch approaching solstice. Therefore, an ice giant mission arriving in the  late 2040s provides the opportunity to sample these systems in different seasonal states compared to those observed by Voyager 2.   

The upper atmosphere science objectives of future studies and new spacecraft mission concepts involve addressing a set of high-level science questions, which could include: 
\begin{enumerate}
\item {\bf Fundamental Parameters} -- What is the current temperature of the upper atmosphere of Neptune? This has only been measured once -- over three decades ago \citep{1989Sci...246.1459B}. Modelling presented in this special issue by {\it Moore et al.} \citep{moore2020} suggests that the upper atmosphere may have significantly cooled since the era of Voyager 2. What mechanism could cause the upper atmosphere of both Uranus and Neptune to cool over time-scales of several decades? What is the relative importance of particle ionisation versus solar EUV ionisation to produce H$_3^+$ at each planet? 

\item {\bf Magnotosphere/Ionosphere Coupling} -- What is the nature of the interaction between the thermosphere, ionosphere, and magnetosphere at Uranus and Neptune? In contrast to Jupiter and Saturn, the ice giants have fainter auroral emissions, indicating smaller particle precipitation fluxes. However, it is unclear how these ionospheres couple with the surrounding space environments in the highly complex magnetic field configurations seen at these planets. These kinds of studies requires a spacecraft in orbit for in-situ measurements of the magnetic field and plasma, combined with multi-spectral remote sensing observations. 

\item {\bf The Energy Crisis} -- What process is responsible for the elevated temperatures observed in the upper atmosphere of the giant planets? Both planets appear to have limited ionosphere-magnetosphere interactions, at least when compared to Jupiter and Saturn, which could severely the limit the amount of energy available for auroral Joule heating. Regardless, the ice giants have very hot upper atmospheres.  The two planets also offer incredible contrast: Uranus has very limited vertical mixing in the stratosphere, which creates a homopause at low altitudes producing a extended upper atmosphere. This is in contrast to Neptune, where strong vertical mixing produce the opposite. 

\item {\bf Comparative Aeronomy} -- The four giant planets display very different upper atmospheres, and the highest contrast is found at the ice giants, driven by their peculiar magnetic field configurations and their very deep (Uranus), and very shallow ionospheres (Neptune). By completing our view of this region at all the planets in our solar system, we can build a comprehensive framework with which to understand generalised thermospheres and ionospheres, which in turn can be applied to the ever growing cohort of exoplanets. 
\end{enumerate}
\section{Summary}

In this brief review we have presented an overview our current understanding of the upper atmosphere of Uranus and Neptune. The Voyager 2 flybys provided the cornerstones for the exploration of these planets, and many of the measurements captured by the spacecraft are the only ones of its kind, e.g. the vertical profiles of temperature of Figure \ref{figec}. This is a clear driver for new robotic missions to the ice giants.

Almost three decades of ground-based near-infrared observations of H$_3^+$ emission from Uranus has produced a unique data set, which has revealed the surprising long-term cooling in the upper atmosphere. The reason for this may be related to seasonally variable Joule heating, or as proposed here, a changing homopause altitude, either with time, or spatially on the planet. H$_3^+$ remains undetected from Neptune. 

A number of facilities provide new ways to explore the ice giants, e.g. JWST and the next generation of ground-based telescopes and instruments. Additionally, as the case builds for a dedicated spacecraft mission to Uranus or Neptune, perhaps in the shape of a large Cassini style mission, the combination of in-situ and remote sensing observations from orbit will without a doubt provide incredibly powerful tools with which to understand the upper atmosphere of these planets.

\vskip6pt

\enlargethispage{20pt}



\aucontribute{HM conceived this review and authored the manuscript. }

\competing{We declare we have no competing interests.}

\funding{HM was supported by a European Research Council Consolidator Grant (under the European Union's Horizon 2020 research and innovation programme, grant agreement No 723890).}

\ack{HM appreciates very helpful discussions with Luke Moore, Leigh Fletcher, and Steve Milan. }



\bibliographystyle{unsrt}

\bibliography{refs}

\end{document}